\documentstyle[12pt,epsfig]{article} 
\begin{document}
\title{Quantum Zeno effect and the detection of gravitomagnetism.}
\author{ A. Camacho
\thanks{email: acamacho@aip.de}~
\thanks{This essay received an ``honorable mention'' in the Annual Essay Competition of the Gravity Research 
Foundation for the year 2000 --- Ed.}\\
Astrophysikalisches Institut Potsdam. \\
An der Sternwarte 16, D--14482 Potsdam, Germany.}

\date{}
\maketitle
\begin{abstract}

In this work we introduce two experimental proposals that could shed some light upon the 
inertial properties of intrinsic spin. In particular we will analyze 
the role that the gravitomagnetic field of the Earth could have on a quantum system with spin $1/2$. We will deduce the expression for Rabi transitions, which depend, explicitly, on the coupling 
between the spin of the quantum system and the gravitomagnetic field of the Earth. Afterwards, the 
continuous measurement of the e\-ner\-gy of the spin $1/2$ system is considered, and an ex\-pre\-ssion for the emerging 
quantum Zeno effect is obtained. Thus, it will be proved that gra\-vi\-to\-mag\-netism, 
in connection with spin $1/2$ systems, could induce not only Rabi transitions but also a quantum Zeno effect. 

\end{abstract}
\newpage
\section{Introduction.}
\bigskip

In more than three--quarters of a century the theory of general relativity (GR) 
has achieved a great experimental triumph. Neverwithstanding, at this point it is also important to comment that all the current direct confirmations 
of GR are confirmations of weak field corrections to the Galilei--Newton mechanics [1]. 
We must also add that one of the most 
important, and yet undetected, predictions of GR is the so called 
gravitomagnetic field [1], sometimes also called Lense--Thirring effect [2], 
which is generated by mass--energy currents. Its measurement would constitute a direct experimental evidence against an absolute inertial frame of reference, and would  
at the same time show the basic role that local inertial frames 
play in nature, i.e., it would be a direct proof that local inertial frames are 
influenced and dragged by mass--energy currents relative to other mass. 

The first efforts in the detection of this gravitomagnetic field 
are quite old [3] and have already included many interesting proposals [4, 5, 6]. 

An additional topic in connection with gravitomagnetism is related to its 
coupling with intrinsic spin, this issue is of fundamental interest since it comprises the 
inertial properties of intrinsic spin. It is noteworthy to comment that this point is under constant analysis [7]. 

In this work we introduce two experimental proposals that could lead to the detection 
of the coupling between intrinsic spin and the gravitomagnetic field. We analyze the 
role that the gravitomagnetic field of the Earth could have 
on a quantum system with spin $1/2$, i.e., our results could allow us to confront 
the effects of mass--energy density currents upon spin. In particular we deduce a Rabi formula, which depends on the coupling 
between the spin of the quantum system and the gravitomagnetic field of the Earth. Afterwards, 
the continuous measurement of the e\-ner\-gy of the spin $1/2$ system is considered, 
and a Zeno effect is obtained. 
\bigskip

\section{Rabi transitions and the gravitomagnetic field.}
\bigskip

Let us consider a spin $1/2$ system immersed in the gravitational field 
of a rotating uncharged, idealized spherical body with mass $M$ and angular momentum $J$. 
In the weak field and slow motion limit the metric, in the Boyer--Lindquist coordinates, 
reads [8]

{\setlength\arraycolsep{2pt}\begin{eqnarray}
ds^2 = -c^2\left( 1 - {2GM\over c^2r}\right)dt^2 + \left( 1 - {2GM\over c^2r}\right)^{-1}dr^2 \nonumber\\
+ r^2\left(d\theta^2 + \sin^2\theta d\phi^2\right) - {4GJ\over c^2r}\sin^2\theta d\phi dt.
\end{eqnarray}}
\bigskip

The gravitomagnetic field in this case is approximately [1]

{\setlength\arraycolsep{2pt}\begin{eqnarray}
\vec {B} = 2{G\over c^2}{\vec {J} - 3(\vec {J}\cdot\hat {x})\hat {x}\over |\vec {x}|^3}.
\end{eqnarray}}
\bigskip

We will assume that the expression that describes the precession of orbital angular 
momentum, immersed, for instance, in the gravitational field of the Earth, can be also used for the 
description of the dynamics in the case of intrinsic spin. This is a natural extension 
of general relativity [7].

Let us now denote the angular momentum of our spherical body by $\vec {J} = J\hat {z}$, 
being $\hat {z}$ the unit vector along the direction of the angular momentum.
Our quantum particle is prepared such that $\vec {S} = S_z\hat {z}$, it has vanishing small 
velocity and acceleration, and it is located on the $z$--axis, with coordinate $Z$. 

There is a 
formal analogy between the weak field and slow motion of the gra\-vi\-to\-magnetic field in general relativity and the magnetic field in electromagnetism [1].
Following this analogy we may write down the interaction Hamiltonian (acting in the two--dimensional spin space of our spin $1/2$ system),
which gives the coupling between $\vec {B}$ and the spin, $\vec {S}$, of our 
particle 

{\setlength\arraycolsep{2pt}\begin{eqnarray}
H = - \vec {S}\cdot\vec {B}.
\end{eqnarray}}
\bigskip

Introducing expression (2) we may rewrite the interaction Hamiltonian as follows

{\setlength\arraycolsep{2pt}\begin{eqnarray}
H = 2{GJ\hbar\over c^2Z^3}\left[|+><+| - |-><-|\right].
\end{eqnarray}}
\bigskip

Here $|+>$ and $|->$ represent the eigenkets of $S_z$. 
Clearly, the introduction of the gravitomagnetic field renders two energy states

{\setlength\arraycolsep{2pt}\begin{eqnarray}
E_{(+)} = 2{GJ\hbar\over c^2Z^3},
\end{eqnarray}}
\bigskip

{\setlength\arraycolsep{2pt}\begin{eqnarray}
E_{(-)} = -2{GJ\hbar\over c^2Z^3},
\end{eqnarray}}
\bigskip

\noindent where $E_{(+)}$ ($E_{(-)}$) is the energy of the spin state $+\hbar/2$ ($-\hbar/2$). 
Let us now define the frequency

{\setlength\arraycolsep{2pt}\begin{eqnarray}
\Omega = \left(E_{(+)} - E_{(-)}\right)/\hbar = 4{GJ\over c^2Z^3}.
\end{eqnarray}}
\bigskip

The present analogy allows us to consider the 
emergence of Rabi transitions [9]. In order to do this let us now introduce a rotating magnetic field, which, at the point where the particle is 
located, has the following form

{\setlength\arraycolsep{2pt}\begin{eqnarray}
\vec {b} = b\left[\cos(wt)\hat {x} + \sin(wt)\hat {y}\right],
\end{eqnarray}}
\bigskip

\noindent where $\hat {x}$ and $\hat {y}$ are two unit vectors perpendicular to the $z$--axis, 
and $b$ is a constant magnetic field.

Under these conditions the total Hamiltonian reads 

{\setlength\arraycolsep{2pt}\begin{eqnarray}
H_T = 2{GJ\hbar\over c^2Z^3}\left[|+><+| - |-><-|\right] \nonumber\\
- {eb\hbar\over 2mc}\left[e^{-iwt}|+><-| + e^{iwt}|-><+|\right].
\end{eqnarray}}
\bigskip

Looking for a solution in the form $|\alpha> = c_{(+)}(t)|+> +~c_{(-)}(t)|->$, we find the 
usual situation [9] (our quantum system has been initially prepared such that $c_{(-)}(0) = 1$ and 
$c_{(+)}(0)= 0$.)

{\setlength\arraycolsep{2pt}\begin{eqnarray}
c_{(-)}(t) = \exp\left[-i{E_{(-)}\over\hbar}t\ + {i\over 2}(w - \Omega)t\right]\left[\cos(\Gamma t) - i{(w - \Omega)\over 2\Gamma}\sin(\Gamma t)\right],
\end{eqnarray}}
\bigskip

{\setlength\arraycolsep{2pt}\begin{eqnarray}
c_{(+)}(t) = i{eb\over 2mc\Gamma}\exp\left[-i{E_{(+)}\over\hbar}t\ - {i\over 2}(w - \Omega)t\right]\sin(\Gamma t).
\end{eqnarray}}
\bigskip

\noindent where $\Gamma = \sqrt{({eb\over 2mc})^2 + {(w - \Omega)^2\over 4}}$.
\bigskip

In this way we find

{\setlength\arraycolsep{2pt}\begin{eqnarray}
{|c_{(-)}(t)|^2\over |c_{(-)}(t)|^2 + |c_{(+)}(t)|^2} = 
\left[ 1 + {({eb\over 2mc\Gamma})^2\sin^2(\Gamma t)\over \cos^2(\Gamma t) + 
{(w - \Omega)^2\over 4\Gamma^2}\sin^2(\Gamma t)}\right]^{-1}.
\end{eqnarray}}
\bigskip

Clearly, the Rabi transitions depend upon the coupling between spin and the gravitomagnetic field. 

{\setlength\arraycolsep{2pt}\begin{eqnarray}
\left(4{GJ\over c^2Z^3} - w\right)^2 = 4\left[\Gamma^2 - \left({eb\over 2mc}\right)^2\right].
\end{eqnarray}}
\bigskip

\section{Quantum Zeno effect and gravitomagnetism.}
\bigskip

Let us now measure, continuously, the energy of our spin $1/2$ system, such that $E$ is the measurement 
output, and that this experiment lasts a time $T$. 
This kind of measuring process can be described by the so called effective Hamiltonian 
formalism [10, 11], which is one of the models that exist in the topic of quantum measurement theory [12]. 
In our case the corresponding effective Hamiltonian reads

{\setlength\arraycolsep{2pt}\begin{eqnarray}
H_{eff} = 2{GJ\hbar\over c^2Z^3}\left[1 + 
i{2\hbar\over T\Delta E^2}\left(E - {GJ\hbar\over c^2Z^3}\right)\right]|+><+|\nonumber\\
- 2{GJ\hbar\over c^2Z^3}\left[1 + 
i{2\hbar\over T\Delta E^2}\left(E + {GJ\hbar\over c^2Z^3}\right)\right]|-><-|\nonumber\\
- {eb\hbar\over 2mc}\left[e^{-iwt}|+><-| + e^{iwt}|-><+|\right] - 
i{E^2\hbar\over T\Delta E^2}\Pi,
\end{eqnarray}}
\bigskip

\noindent where $\Pi$ is the unit operator in the spin space of our particle. Looking for solutions with the form $|\alpha> = c_{(+)}(t)|+> +~c_{(-)}(t)|->$, we deduce

{\setlength\arraycolsep{2pt}\begin{eqnarray}
c_{(-)}(t) = \exp\left[-i{E_{(-)}\over\hbar}t\ - {(E_{(-)} - E)^2\over T\Delta E^2}t
+ i\tilde{\Gamma} t\right]\nonumber\\
\times\left[c_{(-)}(0)\cos(\beta t) - i{c_{(-)}(0)\tilde{\Gamma} + (\gamma/\hbar) c_{(+)}(0)\over \beta}\sin(\beta t)\right],
\end{eqnarray}}
\bigskip

{\setlength\arraycolsep{2pt}\begin{eqnarray}
c_{(+)}(t) = \exp\left[-i{E_{(+)}\over\hbar}t\ - {(E_{(+)} - E)^2\over T\Delta E^2}t
 -i\tilde{\Gamma} t\right]\nonumber\\
\times\left[c_{(+)}(0)\cos(\beta t) + i{c_{(+)}(0)\tilde{\Gamma} - (\gamma/\hbar) c_{(-)}(0)\over \beta}\sin(\beta t)\right],
\end{eqnarray}}
\bigskip

\noindent where $\tilde{\Gamma} = {(w - \Omega)\over 2} + {i\over 2T\Delta E^2}
\left[(E_{(+)} - E)^2 - (E_{(-)} - E)^2\right]$, $\beta^2 = (\gamma/\hbar)^2 + \tilde{\Gamma}^2$, 
and finally $\gamma = -{eb\hbar\over 2mc}$.

Let us now suppose that the measurement output is the energy of the ground state, 
$E_{(-)}$, that we have a resonant perturbation, and that initially only the lowest energy state was populated, in other words, 
$E = E_{(-)}$, $\hbar w = E_{(+)} - E_{(-)}$, and $c_{(-)}(0) = 1$, $c_{(+)}(0) = 0$.
\bigskip

Hence (15) and (16) become

{\setlength\arraycolsep{2pt}\begin{eqnarray}
c_{(-)}(t) = \exp\left[-i{E_{(-)}\over\hbar}t\ - {(E_{(+)} - E_{(-)})^2\over 2T\Delta E^2}t\right]
\left[\cos(\beta t) - i{\tilde{\Gamma}\over \beta}\sin(\beta t)\right],
\end{eqnarray}}
\bigskip

{\setlength\arraycolsep{2pt}\begin{eqnarray}
c_{(+)}(t) = -i{\gamma\over \beta\hbar}\exp\left[-i{E_{(+)}\over\hbar}t\ - {(E_{(+)} - E_{(-)})^2\over 2T\Delta E^2}t\right]\sin(\beta t).
\end{eqnarray}}
\bigskip

Let us now assume that ${(E_{(+)} - E_{(-)})^4\over 4T^2\Delta E^4}> \gamma^2/\hbar^2$, then

{\setlength\arraycolsep{2pt}\begin{eqnarray}
P_{(-)}(t) = 
\left[1 + {\sinh^2({\gamma\over\hbar}\tilde{\Omega}t)\over 
\tilde{\Omega}^2
[\cosh({\gamma\over\hbar}\tilde{\Omega}t) + {\hbar(E_{(+)} - E_{(-)})^2\over 2T\gamma\tilde{\Omega}\Delta E^2}
\sinh({\gamma\over\hbar}\tilde{\Omega}t)]^2}\right]^{-1},
\end{eqnarray}}
\bigskip

\noindent where $\tilde{\Omega} = \sqrt{{\hbar^2(E_{(+)} - E_{(-)})^4\over 4T^2\gamma^2\Delta E^4} - 1}$, $\gamma = -{eb\hbar\over 2mc}$, 
and $P_{(-)}(t) = {|c_{(-)}(t)|^2\over |c_{(-)}(t)|^2 + |c_{(+)}(t)|^2}$. 
\bigskip

In the case $t\rightarrow \infty$ this last expression reduces to 

{\setlength\arraycolsep{2pt}\begin{eqnarray}
P_{(-)}^{(\infty)} = \left[1 + {\left({c^2Z^3\over 4GJ\hbar}\right)^2}{ebT\Delta E^2\over mc}
\left(\sqrt{1 - {({c^2Z^3\over 4GJ\hbar})^4}({ebT\Delta E^2\over mc})^2} -1\right)^{-2}\right]^{-1}.
\end{eqnarray}}
\bigskip

Clearly, Rabi transitions are inhibited, and the asymptotic value that here appears depends 
explicitly upon the coupling between intrinsic spin and the gravitomagnetic field, i.e., $J$ emerges in expression (20). 

At this point it must be commented that the behavior of spin leads, in some cases, 
to the emergence of a non--geometric element in gravity  [13]. 

In this work 
Ahluwalia has considered two different classes of flavor--oscillation clocks. 
The first one comprises the superposition of different mass eigenstates, a\-sso\-ciated to a quantum test particle, 
such that all the terms of the corresponding superposition have the same spin component. 
The second class of flavor--oscillation clocks, contains, at least, two distinct spin 
projections.

If the gravitomagnetic field is absent, then both clocks redshift identically in the 
corresponding gravitational field. Nevertheless, if the source of the gravitational field has a nonvanishing 
angular momentum, then these redshifts do not coincide any more . 
This fact depends not only upon the gravitomagnetic component of the gravitational field, 
but also on the quantum mechanical features of the employed quantum test particle. 
In other words, here a non--geometric element appears when gravitational and quantum mechanical 
phenomena are considered simultaneously. 

Clearly, in the present essay we have a quantum system with spin immersed in a 
nonvanishing gravitomagnetic field. Nevertheless, our case is an eigenstate of the spin 
operator $S_z$, something that in Ahluwalia's second class of flavor--oscillation clocks 
does not happen. This last remark means that our quantum system is closer to Ahluwalia's 
first class of flavor--oscillation clocks than to his second one.
 
Finally, we must add that it is now possible to test, experimentally, the quantum Zeno effect [12], particularly using Penning traps 
to analyze Rabi transitions [14].
\bigskip
\bigskip

\Large{\bf Acknowledgments.}\normalsize
\bigskip

The author would like to thank A. A. Cuevas--Sosa his 
help, and D.-E. Liebscher for the fruitful discussions on the subject. 
It is also a pleasure to thank R. Onofrio for bringing Refs. 6 and 11 to my attention. The hospitality of the Astrophysika\-li\-sches Institut Potsdam is also kindly acknowledged. 
This work was supported by CONACYT Posdoctoral Grant No. 983023.

\end{document}